# Alternative states in plant communities driven by a life-history tradeoff and demographic stochasticity

Niv DeMalach*[1,2] (nivdemalach@gmail.com)

Nadav Shnerb[3]

Tadashi Fukami[1]

[1]Department of Biology, Stanford University, Stanford, California 94305, USA

[2] Institute of Plant Sciences and Genetics in Agriculture, Faculty of Agriculture, Food and Environment, Hebrew University of Jerusalem, Rehovot, Israel

[3] Department of Physics, Bar Ilan University, Ramat Gan, Israel

* Corresponding author

The authors wish to be identified to the reviewers

**ABSTRACT**

Life-history tradeoffs among species are major drivers of community assembly. Most studies investigate how tradeoffs promote deterministic coexistence of species. It remains unclear how tradeoffs may instead promote historically contingent exclusion of species, where species dominance is affected by initial abundances, causing alternative community states. Focusing on the establishment–longevity tradeoff, in which high longevity is associated with low competitive ability during establishment, we study the transient dynamics and equilibrium outcomes of competitive interactions in a simulation model of plant community assembly. We show that, in this model, the establishment–longevity tradeoff is a necessary but not sufficient condition for alternative stable equilibria, which also require low fecundity for both species. An analytical approximation of our simulation model demonstrates that alternative stable equilibria are driven by demographic stochasticity in the number of seeds arriving at each establishment site. This site-scale stochasticity is only affected by fecundity and therefore occurs even in infinitely large communities. In many cases where the establishment-longevity tradeoff does not cause alternative stable equilibria, it still decreases the rate of convergence toward the single equilibrium, resulting in decades of transient dynamics that can appear indistinguishable from alternative stable equilibria in empirical studies.

**INTRODUCTION**

Community assembly is often affected by life-history tradeoffs among species (Pianka 1970, Tilman 1994, Calcagno et al. 2006). Early models investigated how environmental conditions select for different life-history strategies that result in tradeoffs (Pianka 1970, Charnov and Schaffer 1973). Other studies investigated the effects of life-history tradeoffs on species coexistence, with a focus on the competition-colonization tradeoff (Hastings 1980, Tilman 1994, Calcagno et al. 2006). One aspect of life-history tradeoffs that remains poorly understood is their role in driving alternative community states, scenarios where competitive dominance and exclusion of species depend on initial abundances, leading to historical contingency in community assembly. Some theoretical studies have shown that alternative community states are possible under a life-history tradeoff (Kisdi and Geritz 2003, Baudena et al. 2010), but since the primary goal of these studies has been to understand mechanisms of coexistence, the conditions for alternative states remain largely unknown. In addition to advancing curiosity-driven research, this basic knowledge that is currently lacking would help improve the conservation and restoration of degraded systems (Scheffer et al. 2001, Suding et al. 2004, Kadowaki et al. 2018).

In this paper, we propose that alternative community states may be easily caused by a life-history tradeoff. We focus on the establishment–longevity tradeoff, where greater longevity is associated with lower competitive ability during individual establishment. Evidence suggests that traits that increase longevity reduce competitive ability during establishment in many plant communities. For example, comparisons between annual and perennial plants suggest that perennials' longevity is associated with lower specific leaf area, leaf mass fraction, and specific root length (Garnier 1992, Vico et al. 2016). All these traits increase longevity but reduce resource acquisition and therefore competitive ability during establishment relative to annuals (Vaughn and Young 2015). Similarly, the tradeoff arises when grasses have a shorter lifespan than trees but can outcompete their seedlings (Baudena et al. 2010).

Alternative community states arise when each species cannot invade a community dominated by its competitor, i.e., species are at a disadvantage when rare (Ke and Letten 2018). We hypothesized that this outcome could be realized by the establishment-longevity tradeoff. For example, a short-lived species (e.g., annual plants) may be at a disadvantage when rare since its seedlings cannot replace established adults of a long-lived species (Fig 1a). Conversely, the long-lived species may be at a disadvantage when rare if its seedlings are outcompeted by the seedlings of the short-lived species (Fig 1b). In both cases, the species that happen to be initially more abundant will dominate the community permanently, resulting in alternative stable equilibria.

Alternative stable equilibria are not the only mechanism leading to historical contingency in community assembly. When communities that vary in initial species abundances do not converge for a long time, their long-term transient dynamics can also result in a large amount of historical contingency, even when communities would eventually converge into a single equilibrium given enough time (alternative transient states, as opposed to alternative stable states, *sensu* Fukami and Nakajima 2011, 2013). Long-term transient dynamics are a characteristic feature of systems in the vicinity of a point in which a given state becomes unstable (Hastings et al. 2018). Therefore, in many cases, there can be a parameter region of alternative transient states in the periphery of the region where alternative stable states occur (Scheffer et al. 2009).

We developed a simulation model to understand how the establishment–longevity tradeoff might affect community assembly, with special attention to historically contingent outcomes. We investigated possible conditions under which competitive dominance depends on initial abundances, hereafter referred to as priority effects. In this analysis, we sought to disentangle priority effects resulting in alternative stable states from priority effects resulting in long-lasting transient dynamics. As we detail below, our simulation model and its analytical approximation show that the establishment-longevity tradeoff can cause both alternative stable states and alternative transient states. Our analysis also shows that whether alternative stable or transient states occurs depends on other life-history traits, especially fecundity.

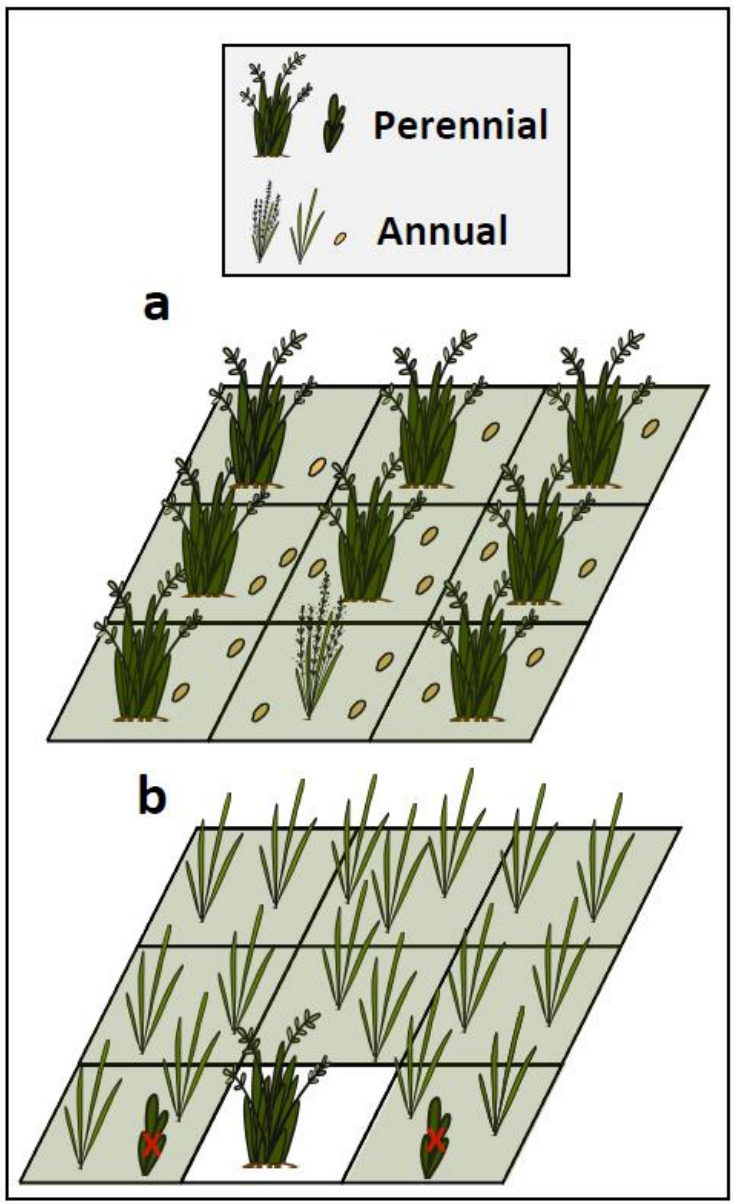

**Figure 1.** A simplified illustration of alternative stable states driven by the establishment-longevity tradeoff. The tradeoff implies that a long-lived species (perennial) is less competitive than a short-live species (annual) during the establishment phase. Alternative stable states arise when both species are at a disadvantage when rare. (a) An annual plant cannot invade a perennial dominated community because its seeds cannot establish in recruitment sites that are occupied by adult perennials. (b) A perennial plant cannot invade an annual-dominated community because its seedlings are outcompeted by annual seedlings.

**METHODS**

We implemented an individual-based, spatially implicit model describing population dynamics of plants where the local site is comprised of many patches (cells), each of which can accommodate only one adult individual (e.g., Mouquet et al. 2002). In this model, plants compete for empty cells but the specific limiting resource is implicit for increasing generality. This approach is well suited for studying life-history trade-offs (Hastings 1980, Calcagno et al. 2006, Gonzalez and Loreau 2009) as it explicitly considers fecundity and mortality. Our simple model describes the dynamics of two grassland species, one annual (semelparous) and one perennial (iteroparous), interpreted as two dominant species or two functional groups with minor differences within each group. As mentioned above, evidence suggests that perennials' advantage of higher longevity often comes at the cost of lower competitive ability at the establishment stage (Bartolome & Gemmill, 1981; Dyer & Rice, 1997; Hamilton, Holzapfel, & Mahall, 1999; Young et al., 2015).

We simulated three processes each year: establishment (competition over recruitment sites), seed production, and mortality. First, at the beginning of the growing season, the establishment of recruits (annuals and perennial seedlings) occurs in patches with no adult perennials (i.e., adult perennials are not affected by annuals or by perennial seedlings). The number of viable seeds of each species arriving at an empty cell is a random Poisson number (a different number for each cell) with the expected value (and SD) equal to the species' mean seed rain ($N$). Mean seed rain is the per-capita net fecundity times the proportion of cells occupied by a given plant species in the previous time step. If at least one seed arrives in an empty cell, then its yearly probability of being occupied by an annual is determined by the following expression: $\frac{C \cdot N_{a(t)}}{N_{p(t)} + C \cdot N_{a(t)}}$, where $N_a$ and $N_p$ represent the mean seed rain of the perennials and annuals and $C$ is a weighting factor describing establishment differences. When $C = 1$, the establishment is completely neutral (i.e., the probability of winning is determined only based on seed density). Higher values of $C$ imply that the annual seeds have higher competitive ability than perennials during the establishment phase i.e., an establishment-longevity tradeoff.

The next process after establishment is seed production by annuals and adult perennials assuming perennial seedlings cannot produce seeds (Mordecai et al. 2015). The net per-capita fecundity for annuals and perennials ($F_a$, $F_p$) is the number of viable seeds per individual of annuals and adult perennials. Biologically, this parameter represents the combined effect of several processes including seed production, seed viability, seed predation, and pathogens. For simplicity, the model does not incorporate seed dormancy.

The last process each year is mortality occurring after seed set (i.e., after the end of the growing season). The yearly survival probability could be viewed as a result of external factors (e.g., disturbance, drought) or as an endogenic demographic trait of the perennials. Survival probability differs between perennial seedlings ($S_s$) and perennial adults ($S_p$). Seedlings that do not die become adults the next year (Mordecai et al. 2015). All annuals die at the end of each growing season.

The fecundity of the annual species ($F_a$) is a free parameter of the model. The fecundity of the perennial species ($F_p$) is a function of annual fecundity and the fecundity coefficient (β): $F_p = \beta \cdot F_a$. This coefficient, ranging from zero to one, determines the strength of fecundity advantage. When $\beta = 1$, the two species have equal fecundity. As $\beta$ decreases, the perennial species suffers from a greater fecundity. This modeling choice allows disentangling the effect of fecundity advantage (determined by $\beta$) and the effect of varying the net fecundity of both species simultaneously (determined by $F_a$).

The parameter space (Table 1) was selected based on empirical data. We aimed to represent a wide range of demographic traits from different systems around the world. The range of each parameter was based on the extremes values of the relevant studies (references below) to increase generality while keeping the results relevant to real plants.

We assumed that annual net fecundity (taking into account both seed production and germination fraction) was in the range of 3-300 (Jakobsson and Eriksson 2000, Dirks et al. 2017, Wainwright et al. 2018). Since preliminary investigations showed that the effect of varying perennial seedling survival was qualitatively similar to varying adult survival, we report only the effects of the latter. We assumed that seedling survival was always 0.3, following Mordecai et al. (2015) and that perennial adult survival was 0.8-0.99, following Fowler (1995), Tuomi et al. (2013), and Mordecai et al. (2015). In additional simulation runs (Appendix S1), we implemented lower survival rates (the entire range, 0-1), which are possible in highly disturbed systems (e.g., arable fields). We did not find any relevant data for parameterizing $C$ or $\beta$ and therefore investigated a wide range (1-30 and 0.3-1, respectively).

For each combination of parameters, we investigated two initial conditions, annuals as residents (90% of the community) and as invaders (10% of the community). We chose this high abundance of invaders and large community size (10,000 patches) for reducing the possibility of stochastic extinction caused by small population size (However this does not means that stochasticity plays a minor role in the dynamic – another source of stochasticity, associated with seeds competition, turns out to be important, see below). Accordingly, preliminary simulations have confirmed that this large community size assures that the same results are obtained in different iterations of the model.

We characterize the dependency on initial conditions (priority effects) in several time steps (individuals were censused after the establishment phase and before mortality occurs). Our operational definition of priority effects is when the annual species has abundance above 50% when started with higher initial abundance but below 50% when starting in low abundance. We fitted a nonlinear regression between simulation length (*t*) and the proportion of the parameter space where priority effects occur (*y*) using an asymptotic function: $y = b_0 + \frac{b_1 \cdot t}{b_2 + t}$. This regression aimed to estimate the equilibrium value of *y* and validating that we have reached equilibrium within the timeframe of our simulations. We found that after 1000 timesteps, all simulations have (asymptotically) reached equilibrium (see Appendix S2 for longer simulations). Therefore, priority effects after 1000 timesteps represent alternative stable states (multiple equilibria), while priority effects in shorter time scales can also represent alternative transient states (slow convergence to a single equilibrium).

To better understand the mechanisms leading to alternative stable states, we have simplified the simulation model by assuming an infinite community size and a single life-stage population structure for perennials (i.e., reproduction starts from the first year and $S_s = S_p$). These simplifications enabled us to reach an analytical approximation that demonstrates general conditions for alternative stable states (Appendix S3).

**Table 1. Parameters of the models**

| Symbol | Description (units) | Value(s) |
|---|---|---|
| $C$ | Competitive difference among seedlings (unitless) | 1,3,30 |
| $F_a$ | Net fecundity of annuals (viable seeds/year) | 3,10,300 |
| $\beta$ | The ratio between annual and perennial fecundities (fraction) | 0.3-1 |
| $S_s$ | Survival probability of perennial seedlings (fraction/year) | 0.3 |
| $S_p$ | Survival probability of adult perennials (fraction/year) | 0.8-0.99 |

## RESULTS

Consistent with previous models of life-history tradeoffs (Charnov and Schaffer 1973, Iwasa and Cohen 1989), our model predicts that dominance by annuals decreases as the survival of adult perennials increases (Fig. 2, Appendix S1), i.e., annuals are favored when perennial adult survival is low. Nonetheless, dominance is also affected by initial frequency under some conditions, leading to either alternative stable states (Fig. 2c-d) or alternative transient states (Fig. 2e-f).

Specifically, alternative stable states (priority effects lasting for 1000 years) occur in the intermediate parameter space between annual dominance and perennial dominance (Fig. 3). We also found that alternative stable states require the establishment-longevity tradeoff (i.e., $C > 1$) and low fecundity for both species (as explained above $F_a$ affects the fecundity of both species simultaneously). Within 1000 years, the less dominant species always reach abundance below 1%, which we interpret as competitive exclusion (preliminary simulations showed that complete exclusion depends on the arbitrary choice of the number of patches).

Priority effects lasting for 50 years occur in a larger region of the parameter space (Fig. 4) compared with equilibrium results (Fig. 3). These findings indicate that in many cases dependency on initial conditions results in long-term transient dynamics rather than alternative stable states. Such alternative transient states do not necessarily require the establishment-longevity tradeoff, but occur more often when this tradeoff is present (Fig 4, compare upper panels vs. middle and lower ones). Overall, the proportion of the parameter space that shows priority effects slowly declined over time from c. 37% after 30 years to c. 16% after 100 years and down to c. 8% at equilibrium, (Fig. 5, S1-S5). We obtained qualitatively similar results when we introduced environmental variability to the model (Fig. S6-S8).

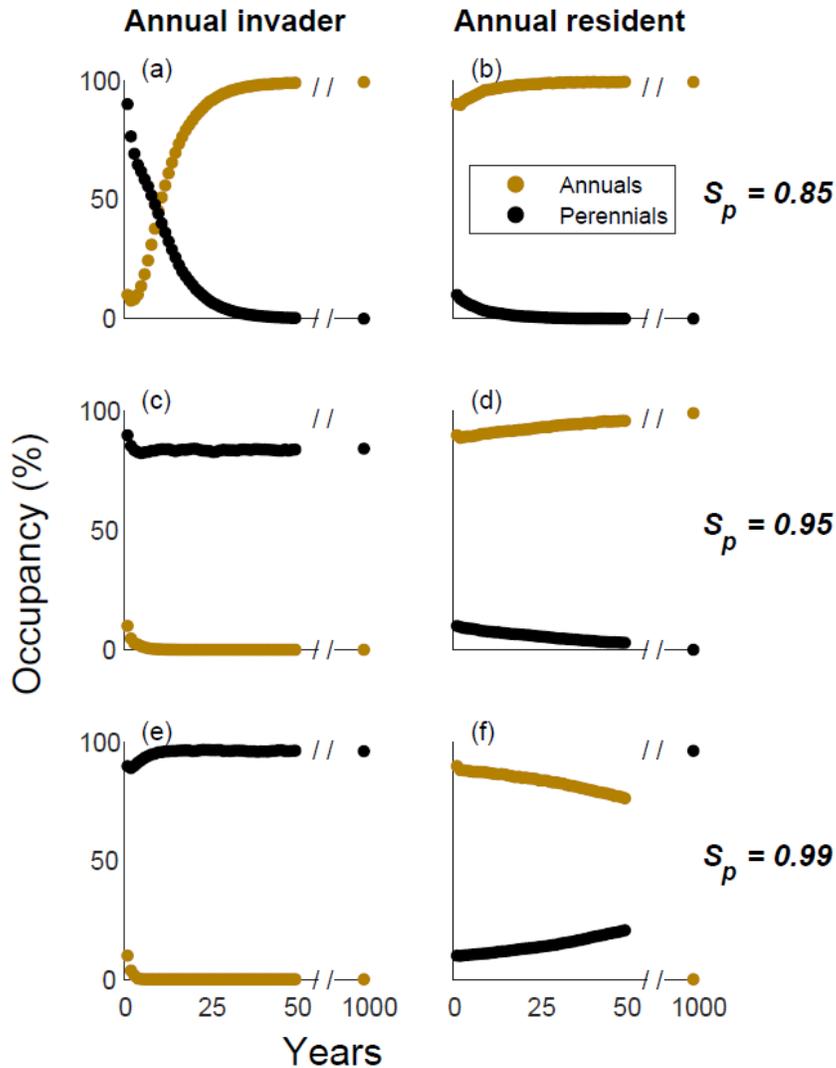

**Figure 2.** Representative examples of population dynamics of annuals (dark yellow lines) and perennials (black lines) as affected by initial conditions and perennial adult survival probability ($S_P$). In the left column, annuals are the minority of the initial community (10%). In the right column, annuals are the majority initially (90%). Low survival probability (0.85) leads to annual dominance (upper panels). Intermediate survival ($S_P$=0.95) leads to alternative stable states where exclusion depends on initial conditions. High levels of adult survival lead to long term transient dynamics where initial conditions affect community dominance for decades although the perennial species dominate in the long-term. Parameter values: $S_s$ =0.3, $F_a$= 5, $\beta$ = 0.5, $C$ =10.

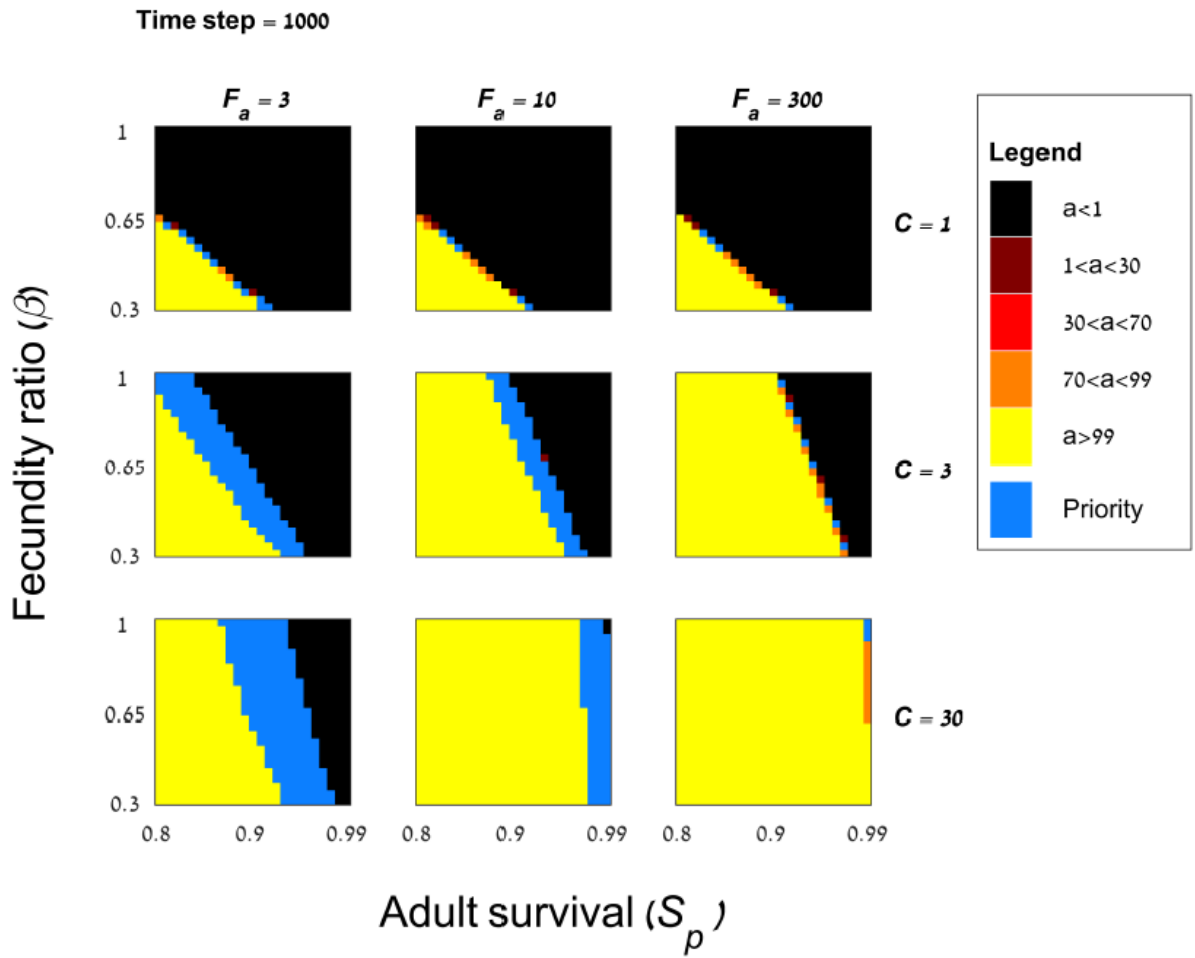

**Figure 3**. The proportion of patches occupied by the annual species ('a' in the legend) after 1000 years (equilibrium results) as affected by competitive differences among seedlings ($C$), annual fecundity ($F_a$), fecundity ratio ($\beta$), and adult survival ($S_p$). The results (of each parameter combination) are the mean proportion of two simulations starting from different initial abundances (10% and 90% of annuals). Black regions represent perennial dominance while yellow regions represent annual dominance. The cases where the dominant species depends on initial conditions (i.e., when annuals comprised more than 50% in one simulation and less than 50% in the other simulation) are categorized as priority effects.

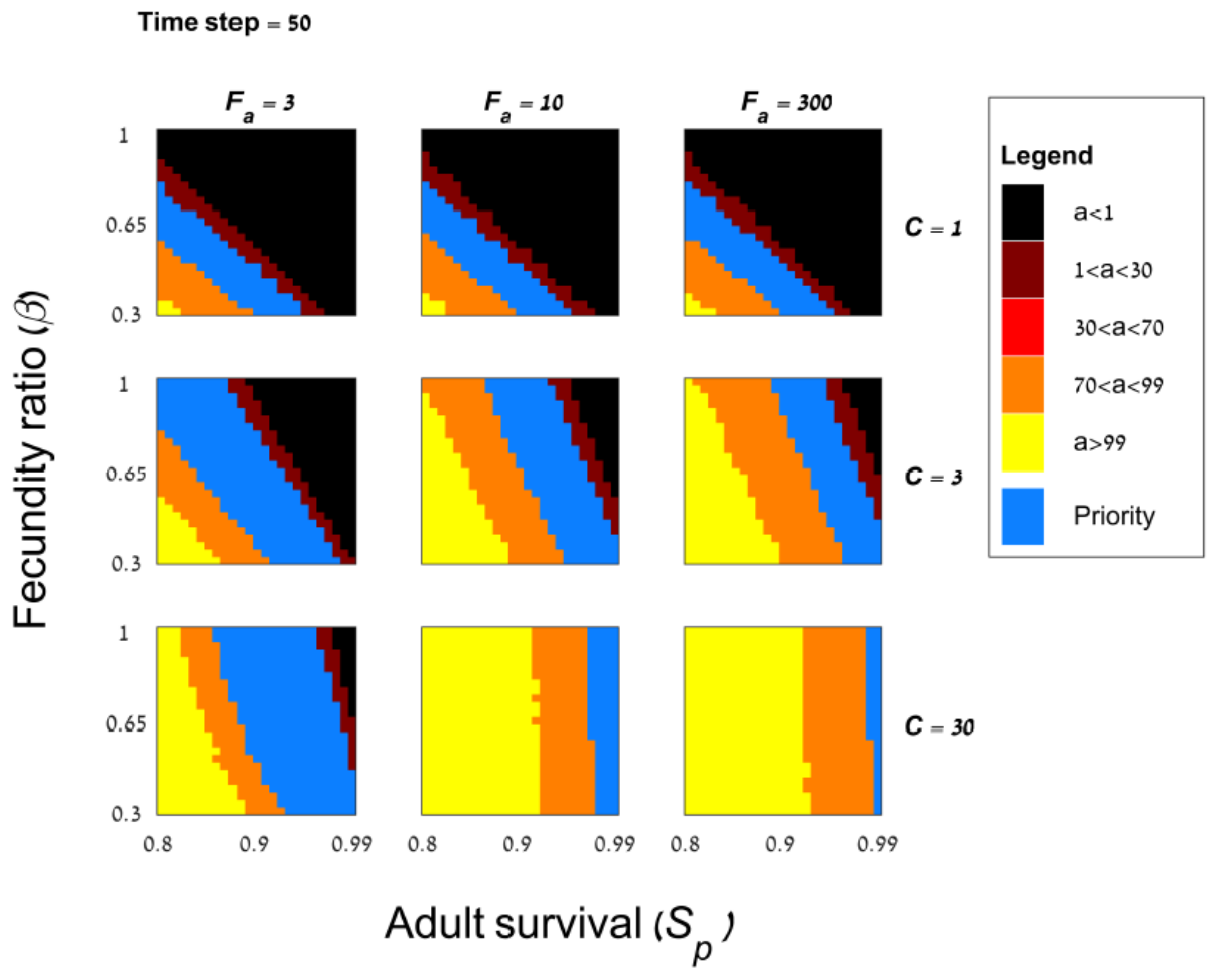

**Figure 4.** The proportion of patches occupied by the annual species ('a' in the legend) after 50 years (transient results). Symbols are as in Fig. 2.

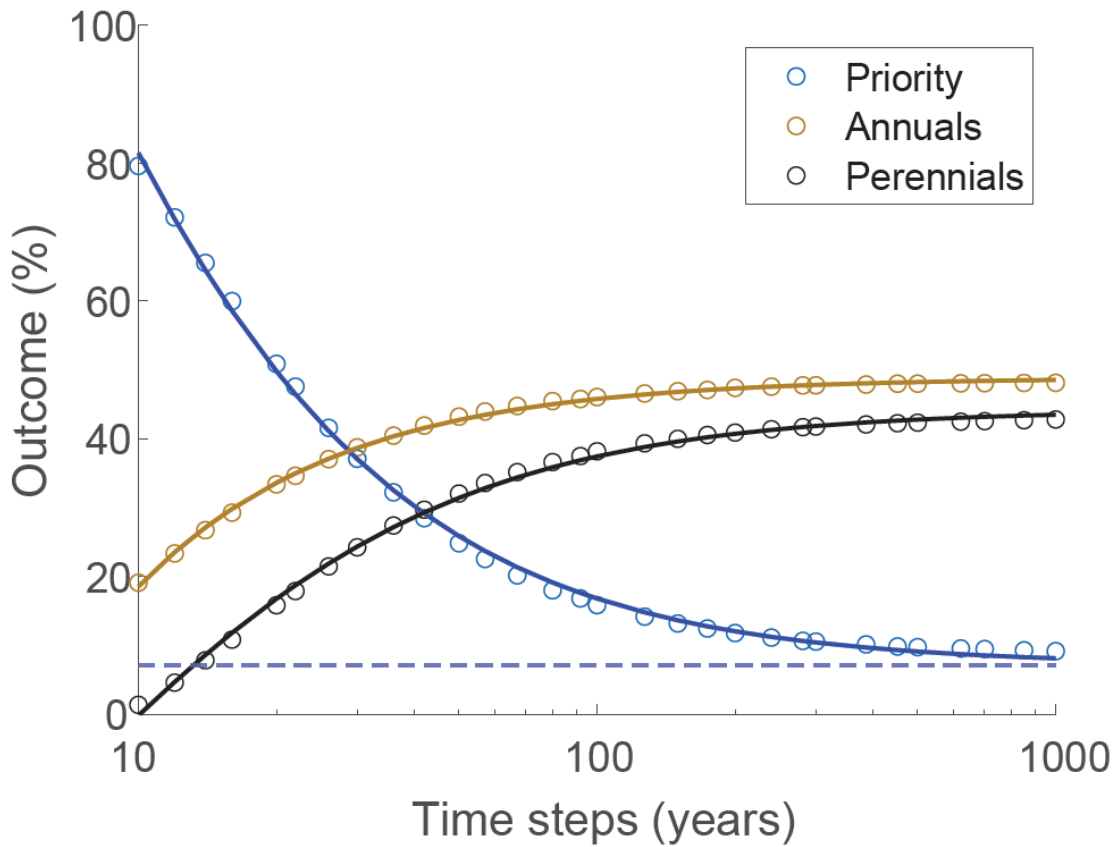

**Figure 5.** The proportion of communities experiencing priority effects (where dominance is determined by initial conditions), annual dominance, and perennial dominance as a function of simulation time (based on the total parameter space shown in Fig 3 and 4). Circles are simulation results and solid lines indicate asymptotic predictions ($y = b_0 + \frac{b_1 \cdot t}{b_2 + t}$). The horizontal blue dashed line represents the asymptotic proportion of priority effects at equilibrium (i.e., alternative stable states; $b_0 + b_1 \approx 8\%$). Note the logarithmic scale of the x-axis. Estimated parameters for priority effects are: $b_0 = 296, b_1 = -288, b_2 = 3$. Estimated parameters for annuals are: $b_0 = -2938, b_1 = 2987, b_2 = 0$. Estimated parameters for perennials are: $b_0 = -72, b_1 = ,116, b_2 = 6$.

The analytical approximation of our simulation model clarifies the role of fecundity ($F_a$) as a necessary condition for alternative stable states (see Appendix S3 for details). Low fecundity increases the demographic stochasticity in seed rain because the number of seeds arriving at each cell is a discrete entity. Therefore, demographic stochasticity in seed rain occurs even when community size is infinite, and its effect strengthens at low fecundities. In other words, under low fecundity, seed rain is low and highly variable regardless of the number of cells.

The probability of establishment of the annual species ($P_a$) in an empty cell that received at least one seed is $P_a = \frac{Cn}{Cn+m}$, where $n$ and $m$ are the numbers of seeds of the annual and the perennial species respectively. In our model, these numbers are drawn from Poisson distributions with means $\bar{n} = F_a x$ and $\bar{m} = F_a \beta(1-x)$, where $x$ and $(1-x)$ are the relative frequencies of the annual and the perennial species. Without stochasticity in seed rain (i.e. assuming $n$ and $m$ in each cell are equal to their expected value), the fecundity ($F_a$) does not affect the dynamic of the model because it is canceled out (appears in both the numerator and denominator). However, when demographic stochasticity is incorporated, fecundity can affect the competitive outcomes because of non-linear averaging (Appendix S3). Under the establishment-longevity tradeoff (when C > 1), demographic stochasticity in seed rain reduces the probability of the establishment of the annual species. Therefore, the annual can be excluded in conditions where it could have won in the deterministic scenario. This negative effect of stochasticity is frequency-dependent, and its strength weakens as annual frequency grows. Alternative stable states occur when the annual has a positive growth only above a threshold frequency and therefore each species is excluded when rare (Fig. 6).

In sum, the analytical approximation of the simulation model demonstrates that the combination of the establishment-longevity tradeoff and low fecundity can produce alternative stable states and that the underlying mechanism is demographic stochasticity in seed rain, which operates against the annual species under low frequency.

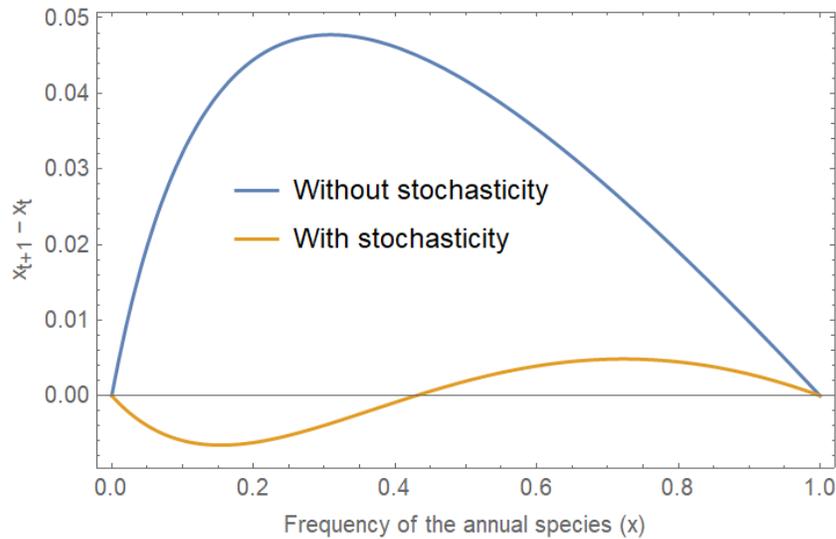

**Fig. 6**: Change in frequency of the annual species ($x_{t+1} - x_t$) as the function of annual frequency ($x$) with and without stochasticity. No demographic stochasticity implies that the seed number of each species in each cell is equal to its expected value rather than a Poisson-distributed variable. Without stochasticity (blue line), the annual species dominate the community and exclude the perennial regardless of initial frequency (change in frequency is never negative) but when stochasticity is incorporated (yellow line), the growth rate is frequency-dependent thereby leading to alternative stable states. Below the threshold frequency (0.45) the annual is excluded by the perennial (i.e., negative change in frequency) and vice versa above it. Results are based on the analytical approximation of the model. Parameter values: $\beta = 0.4, C = 2, F_a = 3$, and $s = 0.7$.

**DISCUSSION**

In our simple model, the establishment–longevity tradeoff sometimes leads to alternative stable states. However, the conditions for alternative stable states are complex because the establishment-longevity tradeoff interacts with other traits, especially fecundity. We have also found that, even when the tradeoff does not produce alternative stable states, it can still decrease the rate of convergence into the dominance of the species with the best strategy, causing alternative transient states. Below, we elaborate on the conditions for alternative states, discuss the limitations of our model, and place our results in the context of existing knowledge on life-history tradeoffs.

**Conditions for alternative stable states and alternative transient states**

In our model, the establishment-longevity tradeoff is a necessary condition for alternative stable states, but not a sufficient one. First, alternative stable states cannot occur when one species has a large demographic advantage over the other. When survival probability of perennials is very high, they dominate regardless of initial abundance and vice versa for annuals when they have high fecundity advantage (i.e., $\beta$ is close to zero). Furthermore, alternative stable states occur only when fecundity of both species is low (as affected by $F_a$). Low fecundity increases demographic stochasticity in seed rain, which reduces the growth rate of the annual species. Under some conditions, the change from negative growth rate to positive growth rate is frequency-dependent such that the annual species is excluded when rare but takes over the community when abundant.

The low net fecundity required for alternative stable states raises the question of how often they occur, especially in the context of annual-perennial interactions, given that many herbaceous plants produce more than 1000 seeds per individual (Jakobsson and Eriksson 2000). This low net fecundity could be possible when considering seed loss caused by pathogens, mechanic decay, and seed predation. The proportion of seeds becoming seedlings is often low, especially in species with high seed output (Ben-Hur, Fragman-Sapir, Singer, & Kadmon, 2012; Muller-Landau, 2010). Moreover, seeds must fall in specific 'recruitment microsites' where the conditions are adequate for germination and early-establishment thereby further reducing the effective number of seeds produced by each individual (Boeken 2018).

In many cases when the establishment-longevity tradeoff cannot produce alternative stable states, the positive feedback created by this tradeoff can still lead to alternative transient states, which may seem similar to alternative stable states in empirical studies when transient dynamics last for decades (Fukami and Nakajima 2011, Hastings et al. 2018).

**Comparison with previous studies**

Unlike this study, most previous studies considered the tradeoff between longevity and competitive advantage as a mechanism of coexistence, not alternative states (Chave, Muller-Landau, & Levin, 2002; Tilman, 1994). This difference is likely to reflect different assumptions about competition. In our model, the better competitor (annual) cannot invade a cell occupied by the inferior competitor (perennial) because seeds cannot replace established adults (replacement competition, *sensu* Yu & Wilson 2001). In contrast, previous models (Chave et al., 2002; Tilman, 1994) have simply assumed that seeds of the best competitor immediately replace established individuals of the inferior competitor (displacement competition, *sensu* Yu & Wilson 2001) and therefore the less competitive species cannot prevent invasion. Evidence suggests that our replacement assumption (where seedlings cannot replace established adults) may be more realistic (Yu and Wilson 2001, Calcagno et al. 2006).

Two previous models have also assumed that seedlings cannot replace adults (Kisdi and Geritz 2003, Baudena et al. 2010). These studies did not investigate alternative states in detail, but both models found bistability in a parameter space where the better competitor had lower longevity. Kisdi & Gertiz's (2003) model also incorporates stochasticity in seed rain and therefore we suspect that the mechanism leading to bistability in their model is the same as in ours. However, Baudena's (2010) model does not incorporate stochasticity, suggesting that, in some cases, the establishment-longevity tradeoff can produce bistability even without demographic stochasticity.

Our results suggest that life-history tradeoff between establishment (during replacement competition) and longevity can be destabilizing (sensu Fukami et al. 2016), i.e., increasing species' performance as their frequency increases thereby leading to alternative stable states. We speculate that such destabilizing life-history tradeoff has not been proposed before, because alternative stable states are often analyzed within the Lotka-Volterra approach where fecundity and survival are lumped into a single parameter, relative growth rate (Ke and Letten 2018). Moreover, in a Lotka-Volterra analog to our model, the competition coefficients would be affected by fecundity ($F_a$) and competitive differences among seedlings ($C$), while in our model they are independent.

**Limitations**

One limitation of our approach is the implicit assumption that species varying in life-history traits have the same adult size (since only one individual can occupy each patch). We used this common modeling approach (Calcagno et al., 2006; Chave et al., 2002; Crawley & May, 1987; Rees & Long, 1992; Tilman, 1994) to facilitate comparison with other models. Nonetheless, since plants vary in size, our model should be viewed as having the extent of cover, rather than the number of individuals, as the unit of abundance.

Another simplifying assumption of our model is the absence of seed dormancy. A similar model of annual-perennial interactions has shown that simply incorporating dormancy into the model (i.e., a constant portion of seeds germinate every year) has modest effects on the model predictions (Rees and Long 1992). However, this model has also suggested that if dormancy is induced by the presence of established perennials (i.e., annual seeds can wait for the right time and replace dead perennials) many predictions could change (Rees and Long 1992). We speculate that such selective germination may further reduce the probability of alternative stable states but may also increase the length of transient dynamics. Furthermore, interactions between seed bank and environmental variability may lead to complex outcomes depending on the specific characteristics of seed dormancy (Brown and Venable 1986, Rees and Long 1992), interactions between seed dormancy and other traits (Venable and Brown 1988), and temporal autocorrelation in environmental conditions (Danino et al. 2016). We found that a simple addition of environmental variability does not affect the model predictions (Appendix S4), but the complex interactions between seed bank and environmental variability in time and space remain to be fully investigated.

Lastly, our model focuses on interactions between two species that represent two functional groups with minor interspecific variability (or two dominant species from each group). High variation within the two groups may lead to deviation from the predicted patterns e.g., in cases where one highly competitive species can outcompete all the rest regardless of initial conditions.

**Demographic stochasticity as a mechanism of alternative stable states**

Historical contingency in community assembly is often viewed as resulting from long-term transient dynamics, alternative stable states, or demographic stochasticity (Chase 2003, Fukami 2015, Hastings et al. 2018). Our model shows that the last two are not independent because demographic stochasticity facilitates alternative stable states. Furthermore, unlike demographic stochasticity of adults that diminish with increasing community size (Gilbert and Levine 2017, Shoemaker et al. 2020), demographic stochasticity in the seed rain is unaffected by community size, only by fecundity. Thus our findings highlight the role of stochasticity in community assembly (Vellend et al. 2014, Shoemaker et al. 2020), especially in grasslands where community size is often large and therefore the role of stochasticity in shaping community structure has been questioned (Vellend 2016).

We believe that non-linear averaging of demographic stochasticity is not a unique attribute of our model and could be relevant to many types of community patterns. For example, the model by Hart et al. (2016) describes how non-linear averaging in demographic traits affects coexistence patterns. While intraspecific demographic variability in that model was interpreted as genetically driven, it could be equally interpreted as an outcome of demographic stochasticity.

**Conclusion**

We have used a simple model to demonstrate how the establishment-longevity tradeoff can lead to alternative stable states and long-term alternative transient dynamics. Although we parameterized the model based on grassland species, the establishment-longevity tradeoff may exist in other types of plants. For example, herbaceous species are often more competitive than seedlings of woody species while the latter often live longer. Therefore, both tree-grass interactions in the Savanna (Baudena et al. 2010) and shrub-herb interactions in Mediterranean systems (Seifan et al. 2010) could be affected by the historical contingency that the establishment-longevity tradeoff causes. Of course, this tradeoff is not the only potential mechanism of alternative community states. We hope that future empirical studies will assess the role of the establishment–longevity tradeoff in comparison with other mechanisms, including fire, plant-soil feedback, and allelopathy (Staver et al. 2011, van der Putten et al. 2013).

2.

**ACKNOWLEDGMENTS**

We thank Eyal Ben-Hur, Inga Dirks, and Ove Eriksson for providing us with raw data from their studies for estimation of the fecundity range of annual and perennial species; Marc Cadotte, Callie Chappell, Po-Ju Ke, Jesse Miller, Erin Mordecai and Suzanne Ou for providing comments; and Callie Chappell for assistance in improving Figure 1. This work was supported by the Rothschild fellowship (N.D.) and the Terman Fellowship of Stanford University (T.F.).

**AUTHORS' CONTRIBUTIONS**

N.D. and T.F. conceived and designed the study; N.D. performed the simulations. N.S. developed and solved the analytical model. N.D. wrote the first draft. All authors significantly contributed to the writing of the manuscript.

**DATA ACCESSIBILITY**

All simulation codes are available on FigShare:

https://figshare.com/s/fb208ec97ca7dd0f250f

**SUPPORTING INFORMATION**

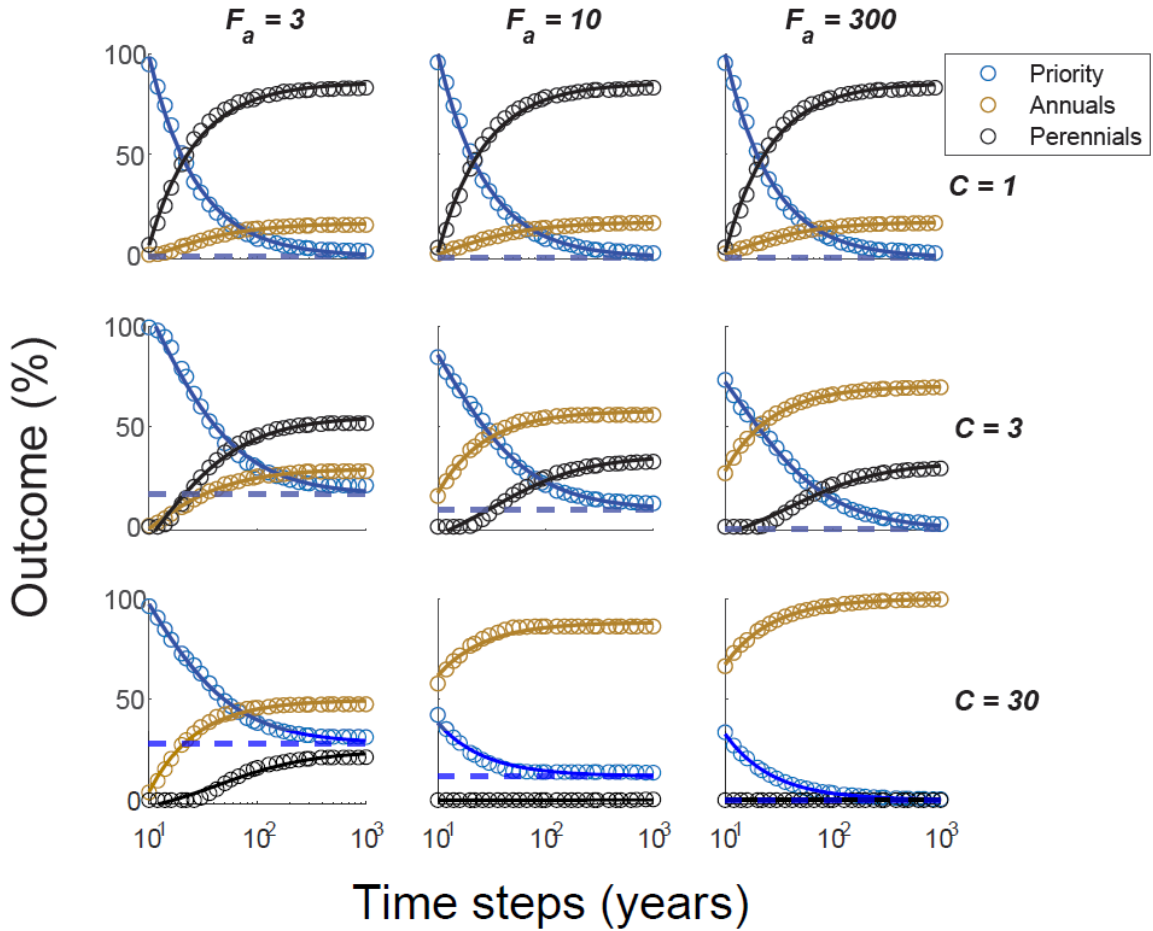

**Fig. S1.** The proportion of communities experiencing priority effects (where dominance is determined by initial conditions), annual dominance, and perennials dominance as a function of simulation time (note the logarithmic scale) as affected by competitive differences *(C)* and annual fecundity ($F_a$). Results are based on combining all levels of fecundity ratio [$\beta$] and adult survival [$S_p$]). Circles are the simulation results and solid lines represent curve fitting of an asymptotic function ($y = b_0 + \frac{b_1 \cdot x}{b_2 + x}$). The dashed blue line represents the equilibrium proportion of priority effects (estimated as $b_0 + b_1$).

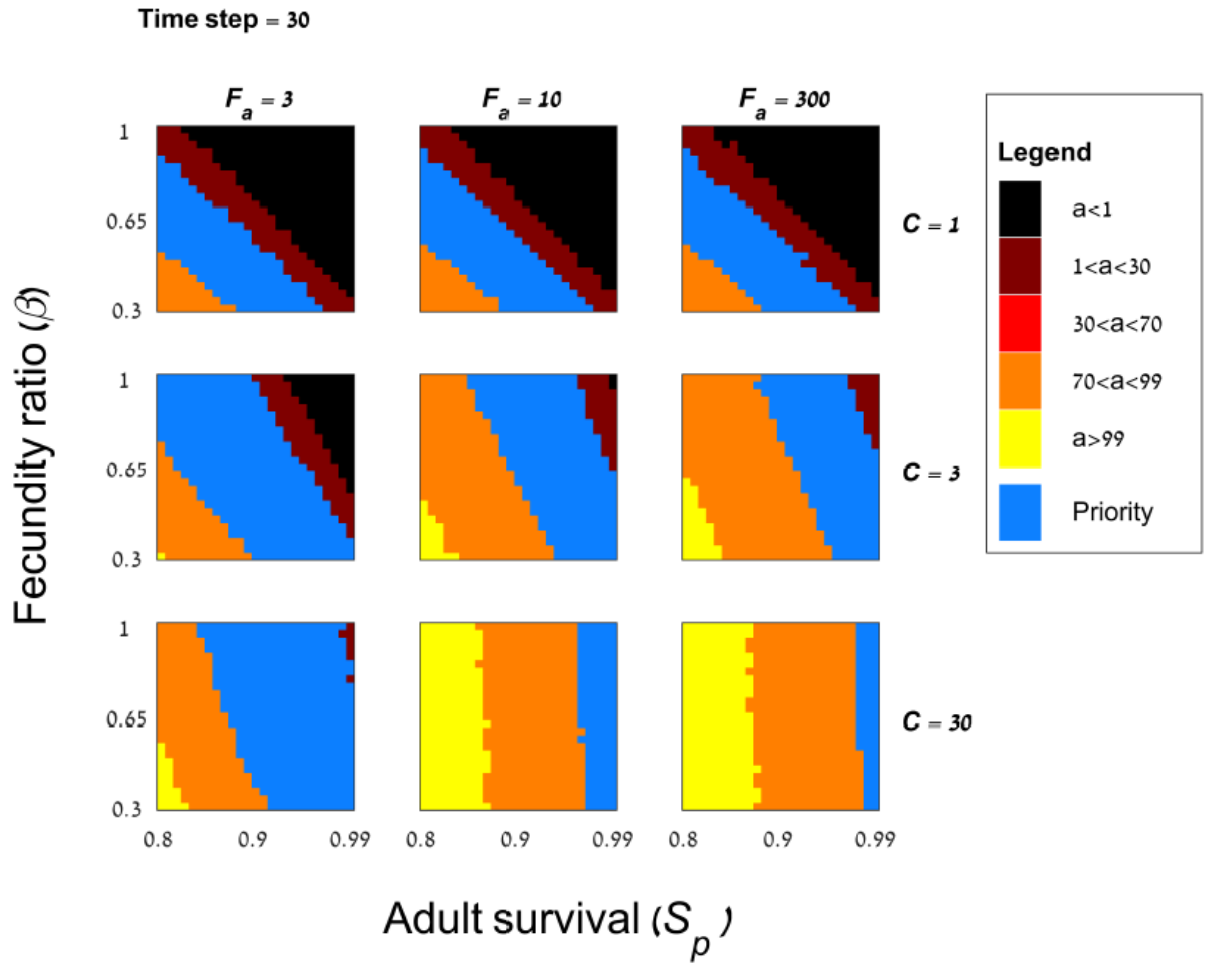

**Fig. S2.** The proportion of patches occupied by the annual species (a) after 30 timesteps (years). Symbols are as in Fig. 2.

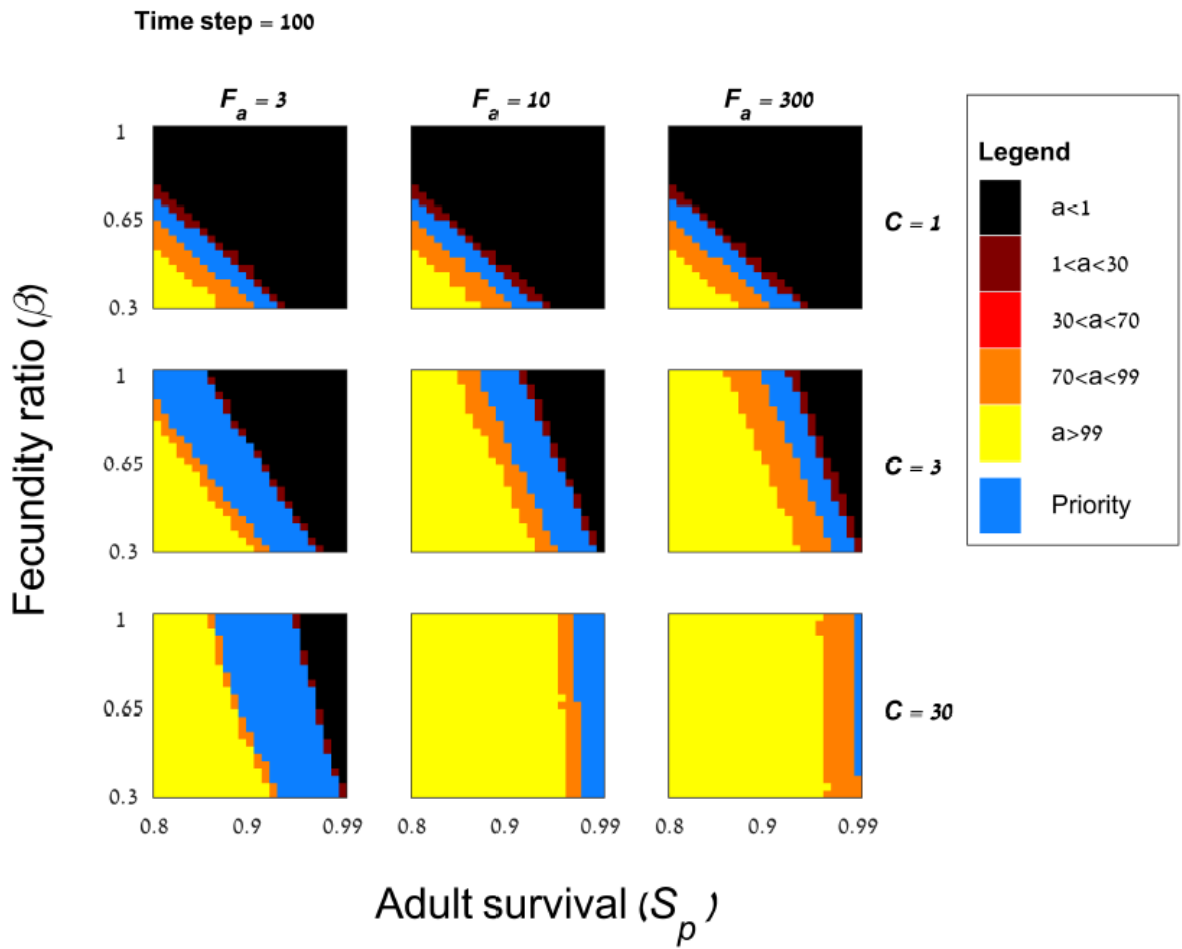

**Fig. S3.** The proportion of patches occupied by the annual species (a) after 100 timesteps (years). Symbols are as in Fig. 2.

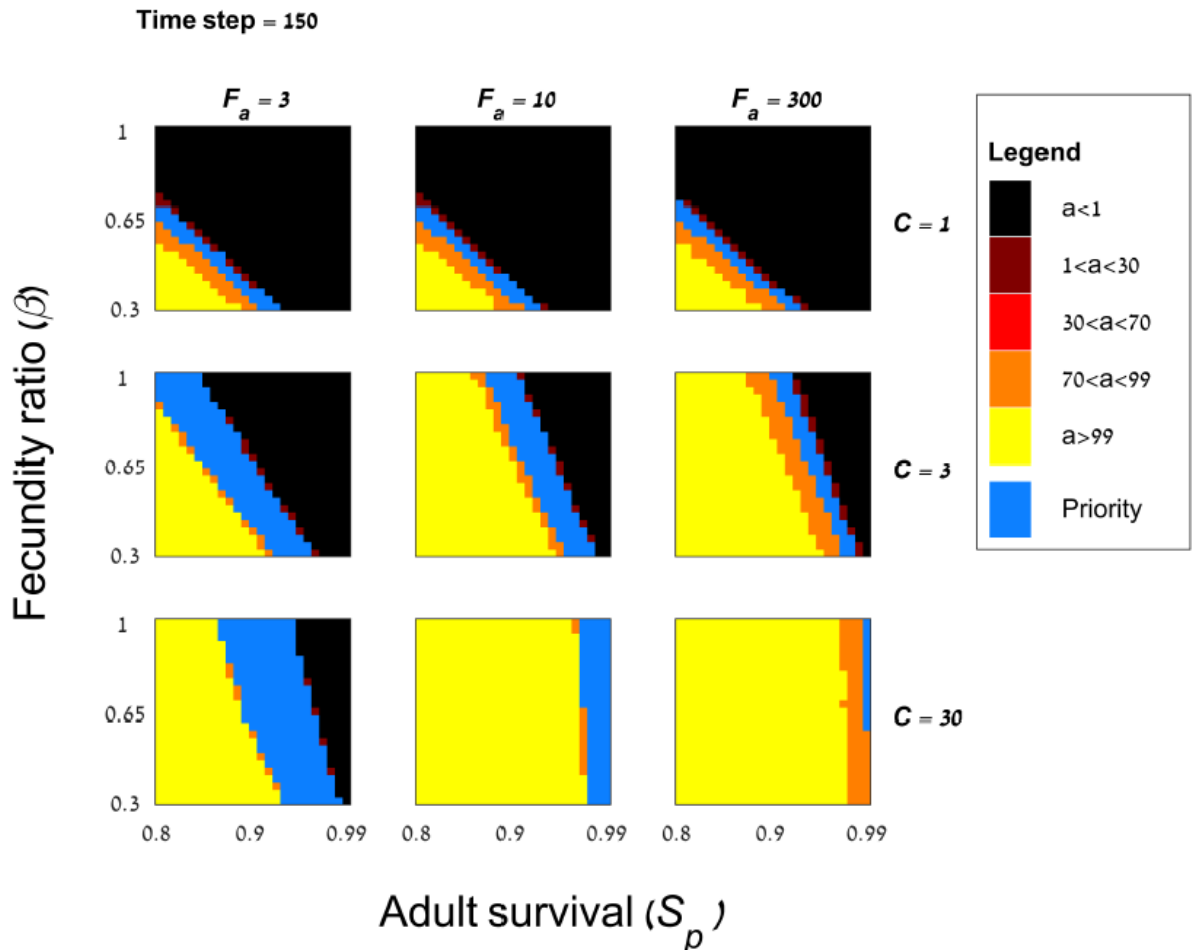

**Fig. S4.** The proportion of patches occupied by the annual species (a) after 150 timesteps (years). Symbols are as in Fig. 2.

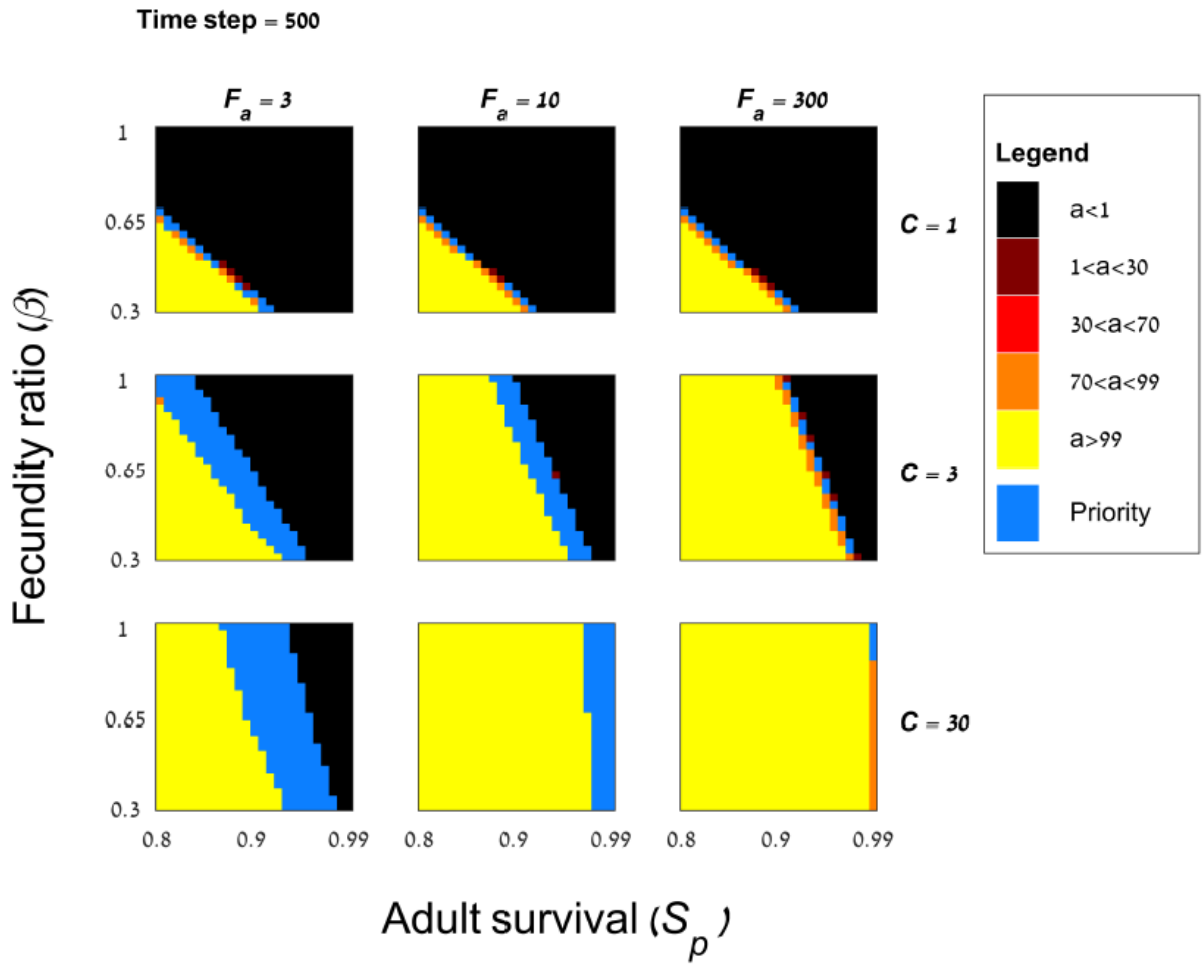

**Fig. S5.** The proportion of patches occupied by the annual species (a) after 500 timesteps (years). Symbols are as in Fig. 2.

# Appendix S1

Here, we investigated a wide range of adult survival probability (0-1 instead of 0.8-1 in the main simulations). While the range of the main simulation was based on empirical data of adult survival in natural systems, a lower survival probability is expected in highly disturbed systems (e.g. arable fields). In accordance with the classical models, we found when the survival probability of adult perennials is low, annuals dominate (Fig S6).

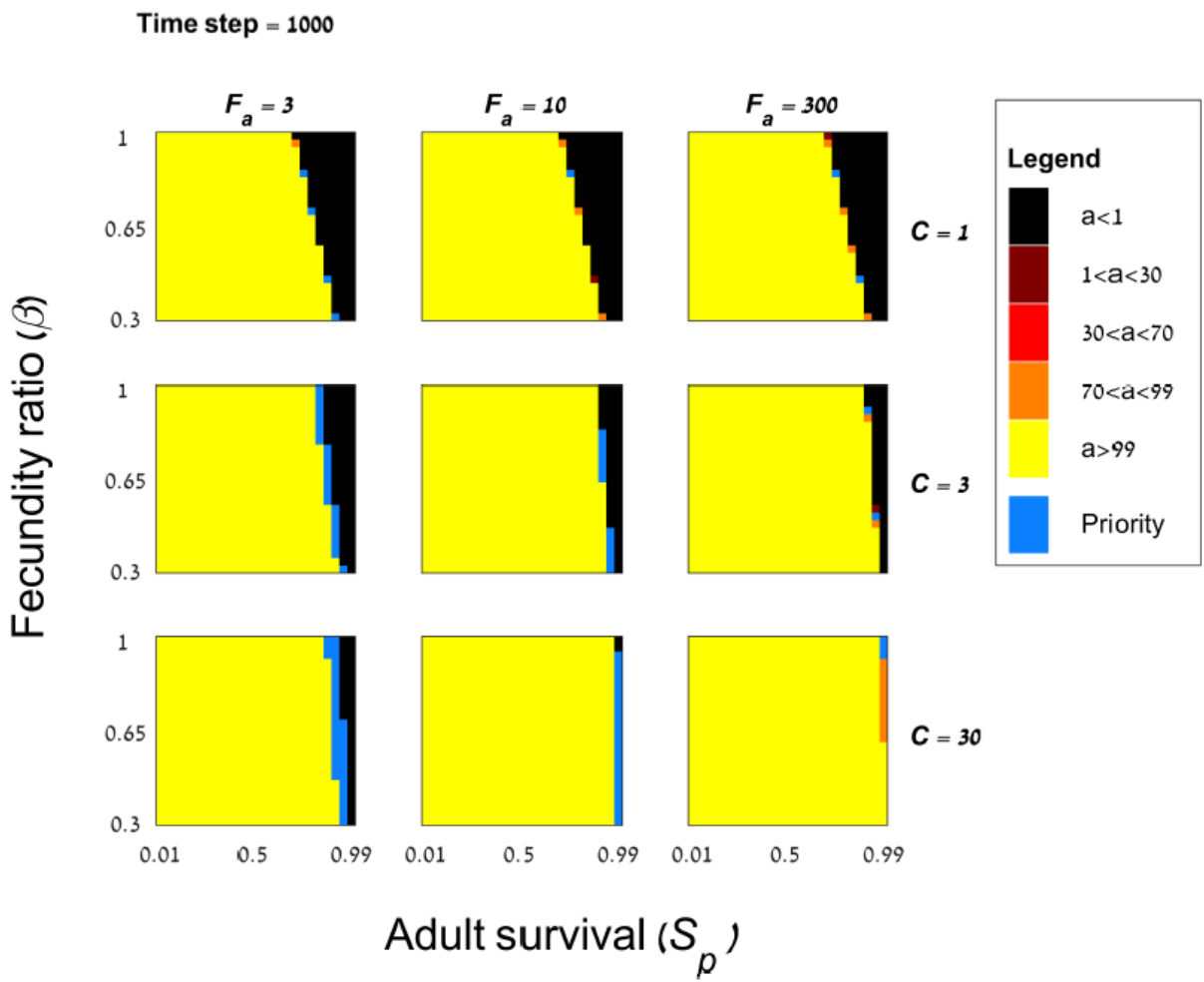

**Fig. S6.** The proportion of patches occupied by the annual species (a) after 1,000 timesteps (years). Symbols are as in Fig. 2.

# Appendix S2

To support our interpretation that priority effects after 1,000 years represent alternative stable states rather than long transient dynamics we have conducted longer simulation of 10,000 timesteps. We investigate the same range of parameters but larger intervals between the values of the continuous parameters. We found that results after 10,000 are very similar to the results after 1,000 (Fig. S7, S8).

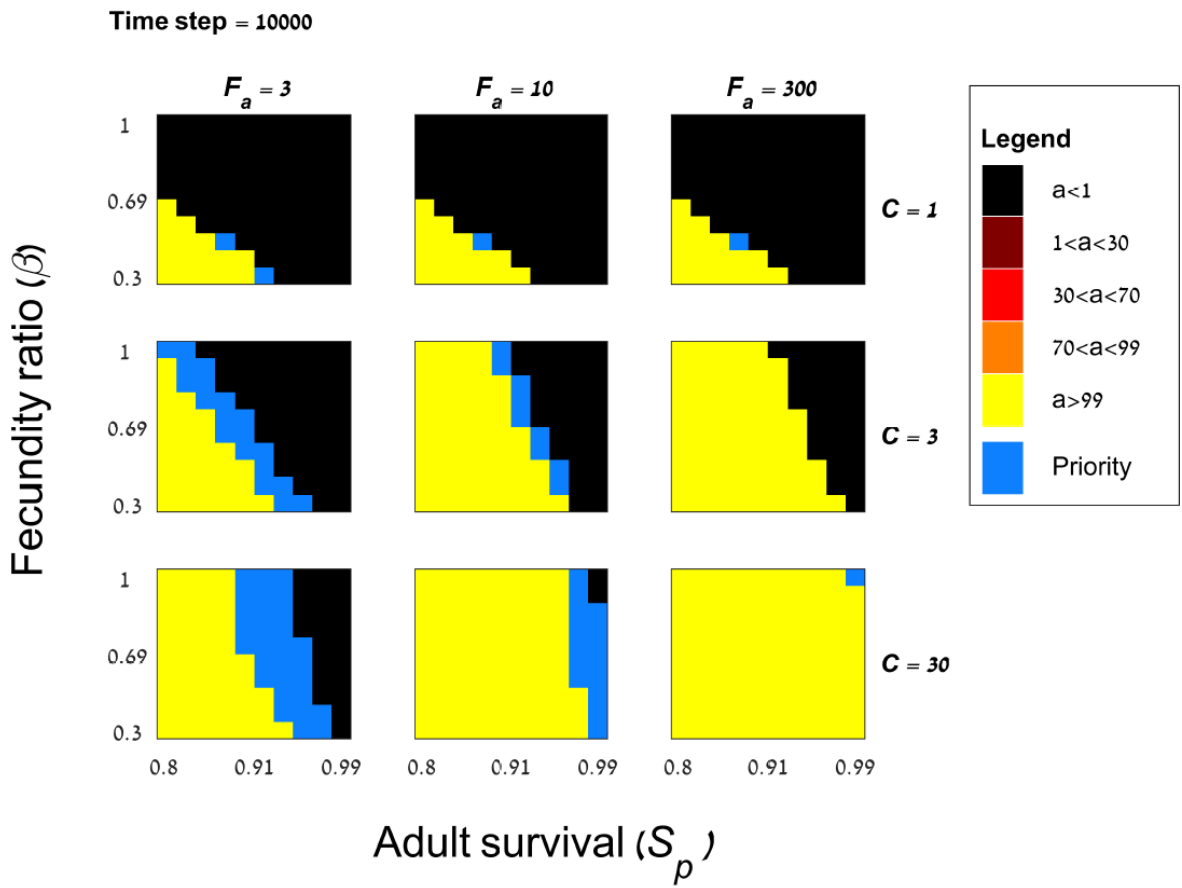

**Fig. S7.** The proportion of patches occupied by the annual species (a) after 10,000 timesteps (years). Symbols are as in Fig. 2.

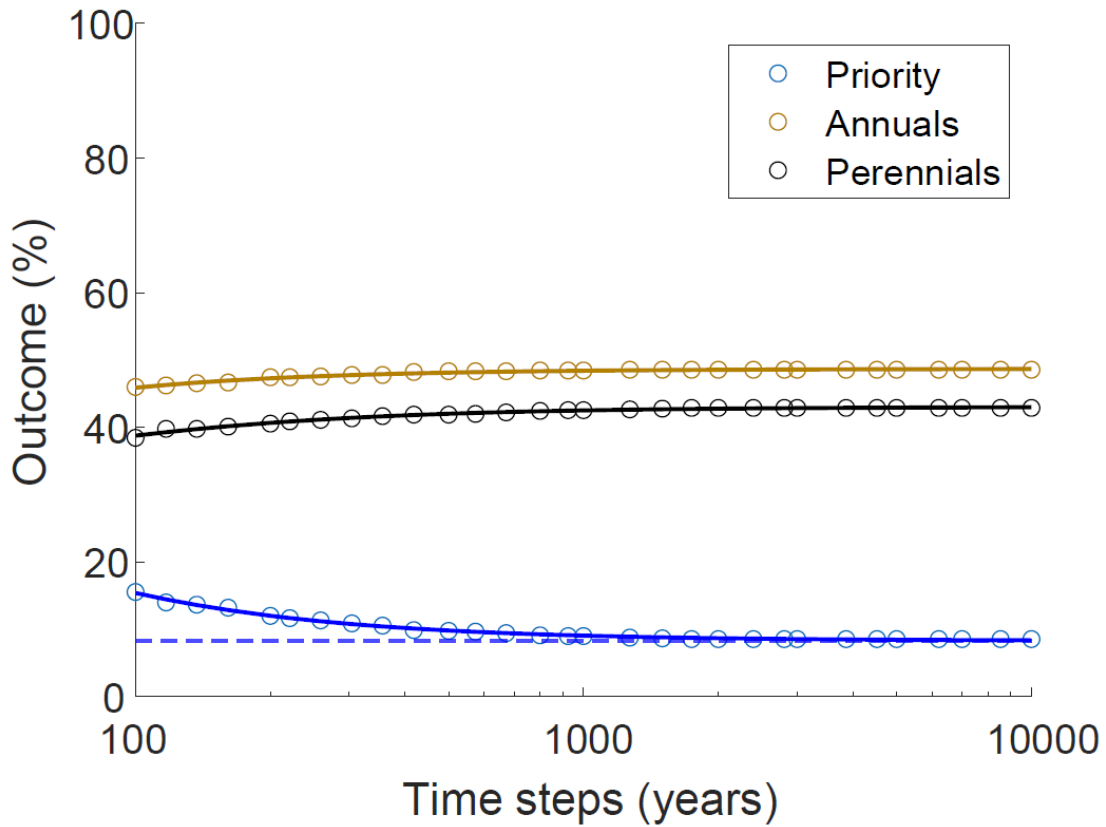

**Figure S8**: The proportion of communities experiencing priority effects (where dominance is determined by initial conditions), annual dominance, and perennial dominance as a function of simulation time (based on the total parameter space shown in Fig Sx). Circles are simulation results and solid lines indicate asymptotic predictions ($y = b_0 + \frac{b_1 \cdot t}{b_2 + t}$). The horizontal blue dashed line represents the asymptotic proportion of priority effects at equilibrium.

# Appendix S3

To better understand the effect of fecundity on model outcomes, we developed an analytical approximation of the simulation model that assumes an infinite number of cells. An infinite community size implies no demographic stochasticity in adults. However, it does not remove stochasticity in seed rain, i.e., the number of seeds in each cell is a random discrete quantity.

Below, we demonstrate that stochasticity in seed rain (hereafter stochasticity) resulting from low fecundity can produce alternative stable states in scenarios when they would not occur otherwise. To that end, we compare deterministic and stochastic cases. In the deterministic case, we assume that the number of seeds of each species in each cell is equal to their expected values, while in the stochastic scenario these values are Poisson-distributed random variables as in our simulations.

We focus on the frequency dynamics of the annual ($x$). For clarity, the frequency of the perennial is approximated to be $1 - x$, assuming that the proportion of empty cells is very small (see details in the section 'proportion of empty cells').

The probability of an annual plant to win an empty cell in the deterministic case is

$$(1)\ P_a^{det} = \frac{CFx}{CFx + \beta F(1-x)}$$

where $F$ is the fecundity of the annual species (subscript $a$ removed for clarity) and $\beta F$ is the fecundity of the perennial. This equation shows that in the deterministic case there is no effect of fecundity on competition outcomes because $F$ canceled out (appears in both the numerator and the denominator).

**Approximation of the stochastic case under high fecundity levels**

Demographic stochasticity in seed rain occurs because the number of seeds in an empty cell must be an integer. The mean number of annual seeds per cell is $\bar{n} = Fx$ and the mean number of perennial seeds is $\bar{m} = \beta F(1-x)$. To calculate the probability of the annual species to take over a given cell in stochastic dynamics, $P_a^{st}$, we average (overline denotes an average) the ratio between two integers, weighted by $C$, where $m$ and $n$ are picked from a Poisson distribution with mean $\bar{n}$ and $\bar{m}$:

(2) $\quad P_a^{st} = \overline{\dfrac{Cn}{Cn+m}}.$

When $F$ is large, one may assume that both random variables $n$ and $m$ are distributed normally around their mean values (because the Poisson distribution converged to the normal distribution as its mean increases), with standard deviations $(\delta n, \delta m)$ that scale like the square root of the mean. Accordingly, when $F$ is large, the standard deviations are much smaller than the corresponding mean values. The second-order Taylor expansion in these small quantities (the mean of $\delta n$ and $\delta m$ vanishes) is:

(3) $\quad P_a^{st} = \overline{\dfrac{C(n+\delta n)}{C(n+\delta n)+(m+\delta m)}}$

$\approx \dfrac{C\bar{n}}{C\bar{n}+\bar{m}} + \dfrac{C\bar{m}\,\overline{\delta n^2} - C^2\bar{n}\,\overline{\delta m^2} + C(C\bar{n}-\bar{m})\,\overline{\delta n \delta m}}{(C\bar{n}+\bar{m})^3}$

$\approx \dfrac{C\bar{n}}{C\bar{n}+\bar{m}} - \dfrac{C(C-1)\beta x(1-x)}{F(Cx+\beta[1-x])^3},$

where the last approximation is based on $\overline{\delta n \delta m} = 0$ because the fluctuations in seed numbers at a given cell are assumed to be uncorrelated between annuals and perennials. The same considerations hold for any other distribution of seed numbers (not necessarily Poisson distribution) as long as the width is much smaller than the mean.

In the stochastic case, the fecundity parameter $F$ does not cancel out, i.e., the probability of the annual to win an empty cell is affected by fecundity. Under the establishment-longevity tradeoff (i.e. when $C > 1$), stochasticity always decreases the probability that the annual species will win. The difference between the stochastic and deterministic cases decreases proportionally to $1/F$ and disappears as $F$ goes to infinity. Taking into account more terms in the Taylor expansion one gets corrections that disappear like $1/F^2$, $1/F^3$ and so on.

**Explicit calculation**

Under the Poisson assumption, the probability to find $n$ annuals and $m$ perennials in a given empty cell is:

(4) $P_{m,n} = e^{-\bar{m}-\bar{n}} \dfrac{\bar{m}^m}{m!} \dfrac{\bar{n}^n}{n!}$.

With probability $P_{m,0}$ there are only annual seeds in that given cell and the annuals win. With probability $P_{0,n}$ there are only perennial seeds and the perennial species wins. A fraction $P_{0,0}$ of the cells remain empty. In all other cases ($m > 0$ and $n > 0$), annual and perennial seeds compete and the chance of the perennial to win is $\dfrac{Cn}{Cn+m}$. Accordingly, the expected chance of the annual to win an empty cell provided that the cell is subject to competition between seeds is

(5) $A = \displaystyle\sum_{\substack{m \geq 1 \\ n \geq 1}} P_{m,n} \dfrac{Cn}{Cn + m}$.

This calculation may be performed explicitly for $C = 2$, and yields

(6) $A = \dfrac{e^{-\frac{(2\bar{m}+\bar{n})^2}{4\bar{m}}} \left(2 e^{\bar{m}+\frac{\bar{n}^2}{4\bar{m}}}(-1+e^{\bar{n}})\sqrt{\bar{m}} + \sqrt{\pi}\mathrm{Erfi}[\frac{\bar{n}}{2\sqrt{\bar{m}}}]\bar{n} - \sqrt{\pi}\mathrm{Erfi}[\frac{2\bar{m}+\bar{n}}{2\sqrt{\bar{m}}}]\bar{n}\right)}{2\sqrt{\bar{m}}}$.

The probability of the annuals to win an empty cell that received at least one seed (not a cell that will remain empty in the next year) is thus,

(7) $P_a^{st} = \dfrac{P_{0,m}+A}{1-P_{0,0}} = \dfrac{A + e^{-\bar{m}}(1-e^{-\bar{n}})}{1 - e^{-\bar{m}-\bar{n}}}$.

As indicated in Fig. S9, this probability is always smaller than $P_a^{det}$ because of non-linear averaging.

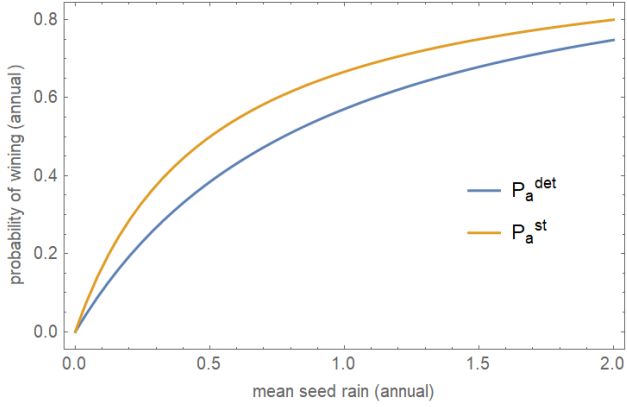

**Fig. S9:** The probability of the annual species to win ($P_a$) as a function of its mean seed rain ($\bar{n}$) with and without demographic stochasticity. The mean seed rain of the perennial is kept equal ($\bar{m} = 1$)

## Demographic stochasticity produces alternative stable states

To demonstrate a scenario where demographic stochasticity in seed rain leads to alternative stable states, we have simplified the model by removing age structure for the perennials, i.e. we assumed that perennials are reproductive from the first year and that their survival rate ($s$) is age-independent (unlike in the simulation model).

The annual species can establish in timestep $t + 1$ only in cells where it was found in the previous time step (the frequency of such cells is $x$) and in cells where perennials have died [with frequency $(1 - s)(1 - x)$]. Therefore, the frequency of the annual species in time step $t + 1$ (after the competition has taken place) is:

(8) $x_{t+1} = P_a \left[ x + (1 - s)(1 - x) \right]$.

When $x_{t+1} - x_t > 0$ the annual species increases in frequency, and vice versa for the perennial species. Since $P_a^{st} < P_a^{det}$ the outcome of the stochastic and deterministic cases can be quantitively different. There are parameter values when the annual always wins (its frequency always grows) under the deterministic map but it wins only above threshold frequency under the stochastic map. The latter scenario implies that the system is characterized by alternative stable states (Fig. 6 in the main text).

**The proportion of empty cells**

Throughout this supplementary we have ignored, for simplicity, the fraction of cells that remain empty after the recruitment step:

$$P_{0,0} = e^{-\bar{m}-\bar{n}}.$$

This assumption was reasonable given that in our simulations, the number of empty cells was extremely small. Moreover, $P_{0,0}$ will be quite small unless the mean total number of seeds per cell is close to zero. To take these empty cells into account one would like to replace $1 - x$, the fraction of perennial in the community, by $1 - P_{0,0} - x$.

# Appendix S4

Here, we investigated whether environmental variability may reduce the dependence on initial conditions. Hence, we assumed that competitive difference ($C$) among recruits varies among years. We chose to vary competitive difference because variation in fecundity always lead to the extinction of the annual species (since there is no seed bank in the model). Competitive differences were a random log-normal variable with a mean of 1, 3, or 30, and SD of 3 for the associated normal distributions (Fig. S10). This approach allowed incorporating variation in time while keeping the mean conditions as in the main simulations (as presented in Fig. 2). We found that the effects of environmental variations were minor (Fig. S11-S12).

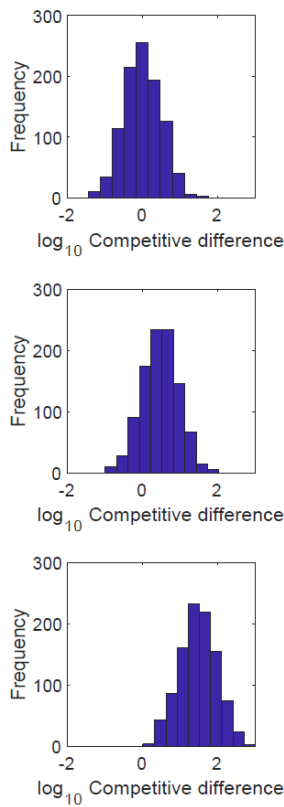

**Fig. S10.** Histograms of competitive differences ($C$) in the simulation (note the logarithmic scale) where environmental variability was incorporated. (a) mean = 1 (b) mean = 3 (c) mean = 30. These values refer to the upper, middle, and lower panels in Fig. S5.

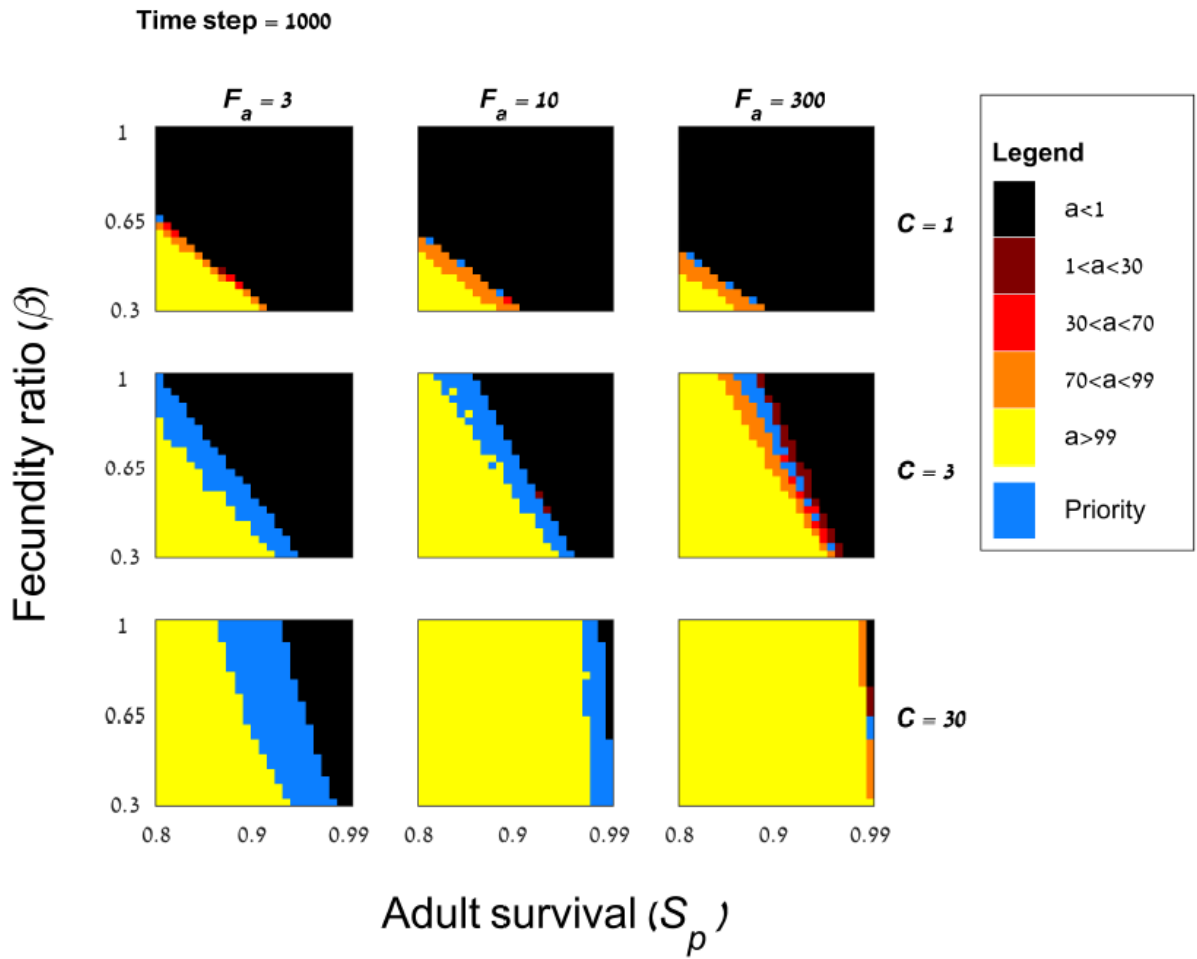

**Fig. S11.** The proportion of patches occupied by the annual species (a) after 1000 timesteps (years) in a temporally variable environment. Competitive differences in each timestep were drawn from log-normal distributions with (geometric) means of 1, 10, and 30. Symbols are as in Fig. 2.

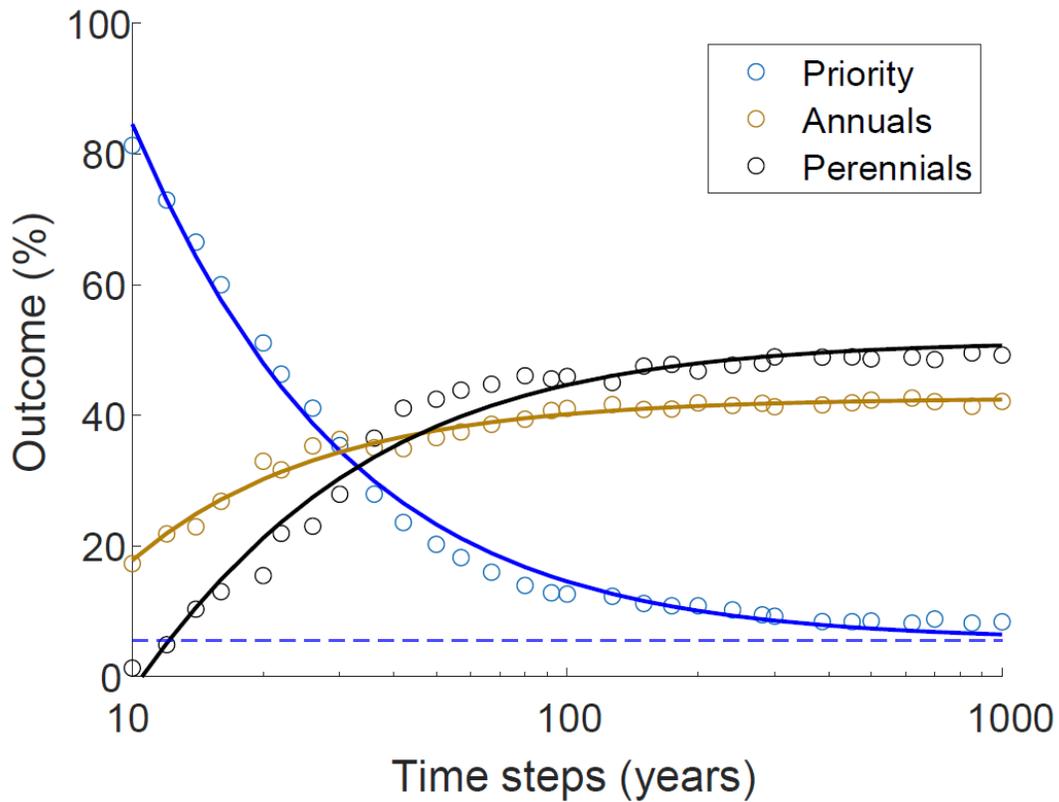

**Fig. S12.** The proportion of communities experiencing priority effects (where dominance is determined by initial conditions), annual dominance, and perennial dominance as a function of simulation time (based on the total parameter space) in a temporally variable environment. Competitive differences in each timestep were drawn from log-normal distributions with (geometric) means of 1, 10, and 30. Symbols are as in Fig. 4. Estimated parameters for priority effects are: $b_0 = 581, b_1 = -275, b_2 = 2$. Estimated parameters for annuals are: $b_0 = -10674, b_1 = 10716, b_2 = 0$. Estimated parameters for perennials are: $b_0 = -182, b_1 = ,233, 0, b_2 = 3$.